\documentclass[twoside]{amsart}
\usepackage{amsmath, amsthm, euscript,epsfig}

\def\Bbb{\mathbb}
\def\cal{\mathcal}
\def\Eu{\EuScript}
 
\def\Ai{\mbox{\rm Ai}}

\def\Re{\mbox{\rm Re}\,}

\def\Im{\mbox{\rm Im}\,}
\def\varkappa{\mbox{\msbm\char'173}}      

\theoremstyle{plain}
\newtheorem{theorem}{Theorem}[section]
\newtheorem{cor}[theorem]{Corollary}

\theoremstyle{definition}

\theoremstyle{remark}
\newtheorem{remark}{Remark}[section]

\numberwithin{equation}{section}

\begin{document}

\title{Quasi-linear Stokes phenomenon for the Painlev\'e first equation}

\author{A.~A.~Kapaev}
\address{St Petersburg Department of Steklov Mathematical Institute,
Fontanka 27, St Petersburg 191011, Russia}
\email{kapaev@pdmi.ras.ru}

\begin{abstract}
Using the Riemann-Hilbert approach, the $\Psi$-function corresponding 
to the solution of the first Painlev\'e equation $y_{xx}=6y^2+x$ with 
the asymptotic behavior $y\sim\pm\sqrt{-x/6}$ as $|x|\to\infty$ is
constructed. The exponentially small jump in the dominant solution 
and the coefficient asymptotics in the power-like expansion to the 
latter are found.
\end{abstract}

\maketitle
\thispagestyle{empty}
\pagestyle{myheadings}
\markboth{A.~A.~Kapaev}
{Quasi-linear Stokes phenomenon for P1}

\section{Introduction}

The Painlev\'e first equation \cite{painleve1}
\begin{equation}\label{P1}\tag{$P_1$}
y_{xx}=6y^2+x,
\end{equation}
is the simplest of the six classical equations of Painlev\'e--Gambier
\cite{ince} and can be derived from any other Painlev\'e equation
using certain scaling reductions \cite{painleve3}. The recent interest
to this equation is due to its significant role in various physical 
models.

For instance, equation $P_1$ describes certain solutions to KdV and 
Bussinesq equations \cite{abl_segur, TSB}, bifurcations in some 
non-integrable nonlinear models \cite{haberman}, continuous limits 
in matrix models of quantum gravity \cite{DS, GM, FIK1, PS}; 
$\Psi$-function associated with $P_1$ appears in $n$-large asymptotics 
of semi-classical orthogonal and bi-orthogonal polynomials 
\cite{FIK2, kapaev:sclim} and thus becomes a primary object in the problem
of Laplacian growth \cite{ABWZ}. 

In context of the string theory, ``physical" solutions of $P_1$ are
distinguished from ``non-physical" ones by  the {\em monotonic} asymptotic 
behavior as $x\to-\infty$ \cite{DS, GM, PS, novikov, FGMR}. There are two 
kinds of such monotonic boundary conditions, i.e. $y(x)\simeq\pm\sqrt{-x/6}$. 
Using elementary perturbation analysis, the solution 
$y(x)=-\sqrt{-x/6}+{\cal O}(x^{-2})$ is unique as being a background to 
a 2-parametric family of oscillating solutions. Solutions approaching 
a positive branch of the square root as $x\to-\infty$, i.e. 
$y(x)\simeq\sqrt{-x/6}$, form a $1$-parametric family parameterized by 
an amplitude of an exponentially small perturbation to a power-like 
dominant solution. These solutions have the asymptotic expansion 
$y(x)=\sqrt{-x/6}\sum_{k=0}^{\infty}a_k(-x)^{-5k/2}+{\cal O}(x^{-\infty})$, 
whose coefficients $a_k$, admit a combinatorial interpretation 
\cite{eynard_zinn-justin, DiFGZJ}.

In the problem of Laplacian growth without surface tension (Hele-Shaw 
problem, quantum Hall effect etc.), the shape of a growing droplet is 
described using certain $\Psi$-function, see \cite{ABWZ}. If the droplet 
develops a cusp singularity, this $\Psi$-function can be approximated by 
a $\Psi$-function associated to the first Painlev\'e transcendent, 
\cite{wiegmann}. Certain ``physical" asymptotic conditions imposed on this 
$\Psi$-function determine the relevant Stokes multipliers $s_k$. In turn, 
these $s_k$ pinch out the monotonic as $x\to-\infty$ Painlev\'e function 
$y(x)\simeq\sqrt{-x/6}$.

Equation $P_1$ has unexpectedly rich asymptotic properties in the complex 
$x$-plane. P.~Boutroux \cite{boutroux} has shown that, generically, 
asymptotics of the Painlev\'e first transcendent as $|x|\to\infty$ is 
described by the modulated Weierstra\ss\ elliptic function whose module 
is a transcendent function of $\arg x$. Furthermore, the module function 
is such that the elliptic asymptotic ansatz degenerates along the directions 
$\arg x=\pi+\frac{2\pi}{5}n$, $n=0,\pm1,\pm2$. P.~Boutroux called the 
corresponding trigonometric expansions {\em ``truncated'' solutions}. Their 
$1$- and $0$-parameter reductions, if they admit analytic continuation into 
one or two neighboring sectors of the complex $x$-plane, were called by 
Boutroux {\em ``bi-truncated''} and {\em ``tri-truncated'' solutions}. All 
bi- and tri-truncated solutions have the algebraic leading order behavior, 
$y(x)\sim\pm\sqrt{-x/6}$, perturbed by exponential terms.

We call a discontinuity in the asymptotic form of an analytic function 
the {\em Stokes phenomenon}. In the case of $P_1$, a jump in 
the phase shift of a modulated elliptic ansatz across the rays 
$\arg x=\pi+\frac{2\pi}{5}n$ is called the {\em nonlinear} Stokes 
phenomenon. For bi- and tri-truncated solutions, a jump in the 
exponentially small perturbation of a dominant solution resembles 
the well known Stokes phenomenon in the linear theory and thus is 
called the {\em quasi-linear} Stokes phenomenon. 

In \cite{BO, HS, hille}, equation $P_1$ was studied further using 
classical tools like the perturbation approach and the method of
nonlinear integral equations. Mainly, these articles discuss the
behavior of the Painlev\'e transcendents on the real line. The recent 
paper \cite{JK} adopts the same approach carefully studying the behavior 
of the tri-truncated solution on the negative part of the real line. 

In \cite{joshi_kruskal, joshi_kruskal2}, the multiple scale analysis 
was applied to $P_1$ (and $P_2$)  to find a precise form of the phase 
shift in the elliptic asymptotic ansatz within complex sectors between 
the indicated rays. In \cite{novikov, FGMR}, the Witham 
averaging method was used to describe the elliptic tail of the monotonic 
at $-\infty$ solution of $P_1$.

The isomonodromy deformation approach to Painlev\'e equations, see 
\cite{jm2, FN, its_nov}, was applied to equation $P_1$ in 
\cite{kapaevP1, kap_kit, kapaev:scl2}. In this way, the asymptotics 
of the Painlev\'e functions is expressed in terms of the Stokes 
multipliers of an associated linear system. Then the equation of 
a monodromy surface yields connection formulas for the asymptotic 
parameters along different directions of the complex $x$-plane. A complete 
description of the nonlinear Stokes phenomenon in $P_1$ is given in 
\cite{kap_kit}. A heuristic description of the quasi-linear Stokes 
phenomenon in $P_1$ can be found in \cite{kapaevP1}. 

Using the Borel transform technique and some assumptions on the analytic 
properties of the relevant Borel transforms, as well as the isomonodromy
deformation approach based on the so-called exact WKB analysis, Y.~Takei 
\cite{takei} has re-derived the latter result (look for more discussion 
in \cite{its_kap}).

In the present paper, we construct the $\Psi$-function associated with the 
monotonic as $|x|\to\infty$ solution of $P_1$ and rigorously describe the 
relevant quasi-linear Stokes phenomenon.  Our main tool is the Riemann-Hilbert 
(RH) problem. We observe that the jump graph for our RH problem can be 
decomposed into a disjoint union of two branches, one of which is responsible 
for the background $\sqrt{-x/6}$ while another one produces the exponentially 
small perturbation of the dominant solution (look \cite{its_kap} for similar 
observation in $P_2$ case). Using the steepest descent approach of Deift and 
Zhou \cite{DZ}, we prove the unique solubility of this problem and compute 
the asymptotics of the Painlev\'e transcendent. 

Applying a rotational symmetry, we prove the existence of five solutions, 
$y_{4n}(x)$, $n=0,\pm1,\pm2$, asymptotic to $\sqrt{e^{-i\pi}x/6}$ as 
$|x|\to\infty$ in the respective overlapping sectors 
$\arg x\in\bigl(
-\frac{\pi}{5}-\frac{4\pi}{5}n,\frac{7\pi}{5}-\frac{4\pi}{5}n)$,
see (\ref{y4n}), (\ref{y4n_ext}), and find the exponentially small
differences $y_{4(n-1)}(x)-y_{4n}(x)$, $n=0,\pm1,\pm2$, see
(\ref{yk_hat_yk}). The latter constitute the quasi-linear Stokes
phenomenon.

A collection of the functions $y_{4n}(x)$, $n=0,\pm1,\pm2$, forms a
piece-wise meromorphic function $\hat y(x)\sim\sqrt{e^{-i\pi}x/6}$ as
$|x|\to\infty$. The moments of this function immediately yield the
asymptotics as $k\to\infty$ for the coefficients $a_k$ (\ref{an_fin}) 
of the $x$-series expansion to the dominant solution (\ref{y+formal}). 

For the first time, the formula for the coefficient asymptotics was 
found in \cite{takei}. The authors of \cite{JK} studied the recurrence 
relations for the same coefficients by direct means and prove a similar 
asymptotic formula modulo a common factor (an advanced version of the 
direct approach to a generalized recurrence relation which contains one 
for $P_1$ as a special case can be found in \cite{HJK}). The exact value 
of this common factor was announced in \cite{JK} with the reference to 
the method of \cite{AK} based on the Borel transform formula. In contrast, 
we do not use the Borel transform technique at any stage of our 
investigation.
 
The paper is organized as follows. In Section~\ref{RH_p1}, we recall
the Lax pair for $P_1$, formulate the relevant RH problem and solve it 
asymptotically. Using the approximate $\Psi$-function, we find the 
asymptotics of the bi- and tri-truncated Painlev\'e transcendents and 
of the relevant Hamiltonian functions. In Section~\ref{as_coeffs}, we 
find the coefficient asymptotics in the power-like expansion to the 
formal solution of $P_1$.

\section{Riemann-Hilbert problem for P1}\label{RH_p1}

Introduce generators of $su(2,{\Bbb C})$,
$\sigma_3=\bigl(\begin{smallmatrix}1&\\&-1\end{smallmatrix}\bigr)$,
$\sigma_+=\bigl(\begin{smallmatrix}&1\\0&\end{smallmatrix}\bigr)$,
$\sigma_-=\bigl(\begin{smallmatrix}&0\\1&\end{smallmatrix}\bigr)$
and the Pauli matrices  $\sigma_1=\sigma_++\sigma_-$ and
$\sigma_2=-i\sigma_++i\sigma_-$  together and consider the system of
matrix differential equations  for $\Psi$, see \cite{garnier1,
garnier3},
\begin{subequations}\label{Lax_pair}
\begin{align}\label{A}
&\frac{\partial\Psi}{\partial\lambda}\Psi^{-1}=A(\lambda,x)=
-z\sigma_3+\bigl(2\lambda^2+2y\lambda+x+2y^2\bigr)\sigma_+
+2(\lambda-y)\sigma_-, \\\label{U} &\frac{\partial\Psi}{\partial
x}\Psi^{-1}=U(\lambda,x)= (\lambda+2y)\sigma_++\sigma_-.
\end{align}
\end{subequations}
Compatibility of (\ref{A}) and (\ref{U}) implies that the coefficients
$z$ and $y$ depend on the deformation parameter $x$ in accord with the
nonlinear differential system
\begin{equation}\label{compatibility}
\begin{cases}
y_x=z,\\ z_x=6y^2+x,
\end{cases}
\end{equation}
which is equivalent to the classical Painlev\'e first equation
\ref{P1}. Following \cite{jm2}, see also \cite{wasow}, linear equation
(\ref{A}) has the only one (irregular) singular point at infinity, and
there exist solutions $\Psi_k(\lambda)$ of (\ref{Lax_pair}) with the
asymptotics
\begin{equation}\label{Psi_k}
\Psi_k(\lambda)=\lambda^{\frac{1}{4}\sigma_3}\tfrac{1}{\sqrt2}
(\sigma_3+\sigma_1) 
\bigl(I-{\Eu H}\sigma_3\lambda^{-1/2}+{\cal O}(\lambda^{-1})\bigr)
e^{\theta(\lambda)\sigma_3},
\quad
\theta(\lambda)=\tfrac{4}{5}\lambda^{5/2}+x\lambda^{1/2},
\end{equation}
as
\begin{equation}\label{Omega_k}
\lambda\to\infty,\quad 
\lambda\in\Omega_k=\bigl\{\lambda\in{\Bbb C}\colon
\arg\lambda\in\bigl(\tfrac{2\pi}{5}\bigl(k-\tfrac{3}{2}\bigr),
\tfrac{2\pi}{5}\bigl(k+\tfrac{1}{2}\bigr)\bigr)\bigr\},\quad 
k\in{\Bbb Z}.
\end{equation}
Solutions $\Psi_k(\lambda)$, $k\in{\Bbb Z}$, are called the {\em
canonical} solutions, while sectors $\Omega_k$ are called the {\em
canonical} sectors. Canonical solutions $\Psi_k(\lambda)$ are uniquely
determined by (\ref{Psi_k})--(\ref{Omega_k}) and solve both equations
(\ref{Lax_pair}). They differ from each other in constant right
matrix multipliers $S_k$ called the {\em Stokes matrices},
\begin{equation}\label{Stokes_matrices}
\Psi_{k+1}(\lambda)=
\Psi_k(\lambda)S_k,\quad S_{2k-1}=
\begin{pmatrix}
1&s_{2k-1}\\ 0&1
\end{pmatrix},
\quad S_{2k}=
\begin{pmatrix}
1&0\\ s_{2k}&0
\end{pmatrix}.
\end{equation}
Observing that all solutions of (\ref{A}) are entire functions, 
thus
\begin{equation}\label{entire}
\Psi_k(e^{2\pi i}\lambda)=\Psi_k(\lambda),
\end{equation}
and using the relation
\begin{equation}\label{cyclic_Psi}
\Psi_{k+5}(e^{2\pi i}\lambda)=\Psi_k(\lambda)i\sigma_1
\end{equation}
which follows from the definition of the canonical solutions and the
asymptotics (\ref{Psi_k}), (\ref{Omega_k}), we readily find the
constraints for the Stokes matrices \cite{kapaevP1},
\begin{equation}\label{Stokes_relations}
S_{k+5}=\sigma_1S_k\sigma_1,\quad S_1S_2S_3S_4S_5=i\sigma_1,
\end{equation}
or, in the scalar form,
\begin{equation}\label{scalar_relations}\tag{\ref{Stokes_relations}$'$}
s_{k+5}=s_k,\quad 1+s_ks_{k+1}=-is_{k+3},\quad k\in{\Bbb Z}.
\end{equation}
Thus, generically, two of the Stokes multipliers $s_k$, $k\in{\Bbb
Z}$, determine all others.

\begin{figure}[hbt]
\begin{center}
\epsfig{file=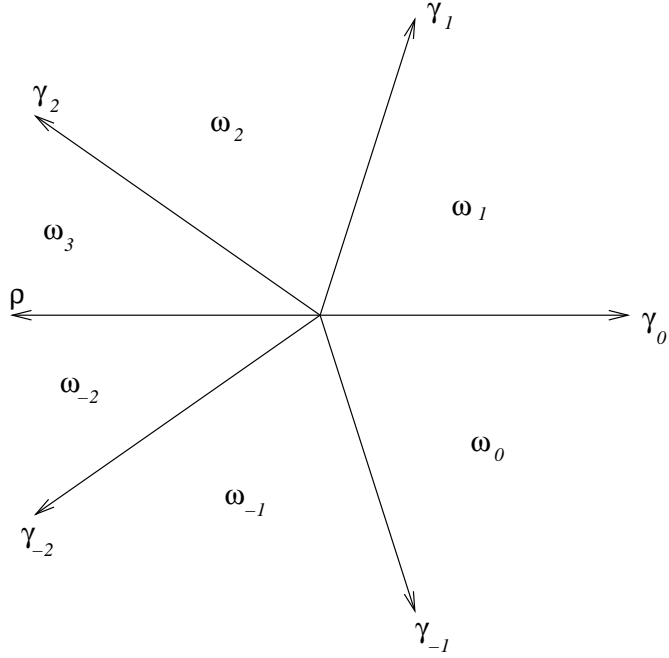}
\end{center}
\caption{The RH problem graph for $P_1$}
\label{f1}
\end{figure}

The inverse monodromy problem consists of reconstruction of
$\Psi_k(\lambda)$ using the known values of the Stokes multipliers
$s_k$. It can be equivalently formulated as a Riemann-Hilbert (RH)
problem. With this aim, introduce the union of rays
$\gamma=\rho\cup\bigl(\cup_{k=1}^5\gamma_{k-3}\bigr)$, where
$\gamma_k=\{\lambda\in{\Bbb C}\colon\ 
\arg\lambda=\frac{2\pi}{5}k\}$,
$k=-2,-1,0,1,2$, and 
$\rho=\{\lambda\in{\Bbb C}\colon\arg\lambda=\pi\}$, all oriented 
toward infinity. Denote the sectors between the rays $\rho$ and 
$\gamma_{-2}$ by $\omega_{-2}$,
between $\gamma_{k-1}$ and $\gamma_k$, $k=-1,0,1,2$, by $\omega_k$,
and between $\gamma_2$ and $\rho$ by $\omega_3$. All the sectors
$\omega_k$ are in one-to-one correspondence with the canonical sectors
$\Omega_k$ (\ref{Omega_k}), see Figure~\ref{f1}.

Let each of the sectors $\omega_k$, $k=-2,-1,\dots,3$, be a domain for
a holomorphic $2\times2$ matrix function $\Psi_k(\lambda)$. Denote the
collection of $\Psi_k(\lambda)$ by $\Psi(\lambda)$,
\begin{equation}\label{Psi_collect}
\Psi(\lambda)\bigr|_{\lambda\in\omega_k}=\Psi_k(\lambda).
\end{equation}
Let $\Psi_+(\lambda)$ and $\Psi_-(\lambda)$ be the limits of
$\Psi(\lambda)$ on $\gamma$ to the left and to the right,
respectively.

Let $\theta(\lambda)=\tfrac{4}{5}\lambda^{5/2}+x\lambda^{1/2}$ be
defined on the complex $\lambda$-plane cut along the negative part of
the real axis. The RH problem we talk about is the following one:
\begin{enumerate}
\item 
Find a piece-wise holomorphic $2\times2$ matrix function
$\Psi(\lambda)$ such that
\begin{equation}\label{p28}
\lim_{\lambda\to\infty}\lambda^{1/2}\Bigl(
\tfrac{1}{\sqrt2}(\sigma_3+\sigma_1)\lambda^{-\frac{1}{4}\sigma_3}
\Psi(\lambda)e^{-\theta\sigma_3}-I\Bigr)
\quad\text{exists and is diagonal};
\end{equation}
\item
on the contour $\gamma$, the jump condition holds
\begin{equation}\label{p29}
\Psi_+(\lambda)=\Psi_-(\lambda)S(\lambda),
\end{equation}
where the piece-wise constant matrix $S(\lambda)$ is given by
equations:\hfill\par
\noindent
\begin{subequations}\label{jump_matrices}
\begin{align}\label{p23}
&S(\lambda)\bigr|_{\gamma_k}=S_k,\quad
S_{2k-1}=I+s_{2k-1}\sigma_+,\quad S_{2k}=I+s_{2k}\sigma_-,
\\\label{p23'} &S(\lambda)\bigr|_{\rho}=-i\sigma_1,
\end{align}
\end{subequations}
with the constants $s_k$ satisfying the constraints
(\ref{scalar_relations});
\end{enumerate}

Because $\Psi(\lambda)$ satisfies the asymptotic condition
\begin{multline}\label{Y_expansion}
Y(\lambda):=
\tfrac{1}{\sqrt2}(\sigma_3+\sigma_1)\lambda^{-\frac{1}{4}\sigma_3}
\Psi(\lambda)e^{-\theta\sigma_3}= 
\\ 
=\begin{pmatrix} 
1-\frac{\Eu H}{\lambda^{1/2}}+\frac{{\Eu H}^2}{2\lambda} 
+{\cal O}(\lambda^{-3/2})& \frac{y}{2\lambda}+{\cal O}(\lambda^{-3/2})\\
\frac{y}{2\lambda}+{\cal O}(\lambda^{-3/2})& 
1+\frac{\Eu H}{\lambda^{1/2}}+\frac{{\Eu H}^2}{2\lambda} 
+{\cal O}(\lambda^{-3/2})
\end{pmatrix},\quad
\lambda\to\infty,
\end{multline}
where
\begin{equation}\label{Ham_def}
{\Eu H}=\tfrac{1}{2}z^2-2y^3-xy,
\end{equation}
the solution $y(x)$ of the Painlev\'e equation can be found from the
``residue" of $Y(\lambda)$ at infinity,
\begin{equation}\label{y_from_Y}
y=2\lim_{\lambda\to\infty}\lambda Y_{12}(\lambda)=
2\lim_{\lambda\to\infty}\lambda Y_{21}(\lambda).
\end{equation}

\begin{remark}\label{rem1}
It is easy to see that ${\Eu H}$ is nothing but the Hamiltonian for
the Painlev\'e first equation with the canonical variables $q=y$ and
$p=z$.
\end{remark}

Equation (\ref{y_from_Y}) specifies the Painlev\'e transcendent as a
function $y=f(x,\{s_k\})$ of the deformation parameter $x$ and of the
Stokes multipliers $s_k$. Using the solution $y=f(x,\{s_k\})$ and the
symmetries of the Stokes multipliers described in \cite{kapaevP1}, we
obtain further solutions of \ref{P1}:
\begin{subequations}\label{P_symmetries}
\begin{align}\label{P_symmetries_bar}
&y=\overline{f(\bar x,\{-\overline{s_{-k}}\})},
\\\label{P_symmetries_rotate}
&y=e^{i\frac{4\pi}{5}n}f(e^{i\frac{2\pi}{5}n}x,\{s_{k+2n}\}),\quad
n\in{\Bbb Z},
\end{align}
\end{subequations}
where the bar means the complex conjugation.

By technical reason, to find the asymptotics of $y(x)$, we use below
not $Y(\lambda)$ but related auxiliary functions $\chi(\lambda)$ and
$X(\lambda)$ with expansions (\ref{chi_expansion}) and
(\ref{X_expansion}), respectively. The latter involve differences
$y-\hat y(x)$, $\hat y(x)$ are known, which can be estimated using
singular integral equations with contracting operators.

\subsection{Asymptotic solution for $s_0=0$}

Let us consider the RH problem above where $s_0=0$ assuming that
$|x|\to\infty$ within the sector $\arg
x\in[\frac{3\pi}{5},\pi]$. Equations (\ref{p23}) imply that
$\Psi(\lambda)$ has no jump across the ray $\gamma_0=\{\lambda\in{\Bbb
C}\colon\arg\lambda=0\}$. The constraints (\ref{scalar_relations})
reduce to the following system of equations,
\begin{equation}\label{s13_relation}
s_{-2}=s_2=s_{-1}+s_1=i.
\end{equation}

Our first step in the RH problem analysis consists of introduction an
auxiliary $g$-function,
\begin{equation}\label{g_function}
g(\lambda)= \tfrac{4}{5}(\lambda+2\lambda_0)^{5/2}
-4\lambda_0(\lambda+2\lambda_0)^{3/2},\quad
\lambda_0=\sqrt{e^{-i\pi}x/6},
\end{equation}
defined on the complex $\lambda$-plane cut along the ray
$(-\infty,-2\lambda_0]$. The asymptotics of $g$-function as
$\lambda\to\infty$ coincides with the canonical one,
\begin{equation}\label{g_as}
g(\lambda)=\tfrac{4}{5}\lambda^{5/2}-6\lambda_0^2\lambda^{1/2}
-4\lambda_0^3\lambda^{-1/2}+{\cal O}(\lambda^{-3/2})
=\tfrac{4}{5}\lambda^{5/2}+x\lambda^{1/2} +{\cal O}(\lambda^{-1/2}).
\end{equation}

Let us formulate an equivalent RH problem for the piece-wise
holomorphic function $Z(\lambda)$,
\begin{equation}\label{Z_def}
Z(\lambda)=Y(\lambda)e^{(\theta(\lambda)-g(\lambda))\sigma_3}=
\tfrac{1}{\sqrt2}(\sigma_3+\sigma_1)\lambda^{-\frac{1}{4}\sigma_3}
\Psi(\lambda)e^{-g(\lambda)\sigma_3},
\end{equation}
\begin{align}\label{Z_RH}
i)\quad &Z(\lambda)\to I\quad\text{as}\quad \lambda\to\infty;\notag \\
ii)\quad &Z_+(\lambda)=Z_-(\lambda)G(\lambda),\quad
G(\lambda)=e^{g\sigma_3}S(\lambda)e^{-g\sigma_3},\quad
\lambda\in\gamma_k, \\
&Z_+(\lambda)=\sigma_1Z_-(\lambda)\sigma_1,\quad \lambda\in\rho.\notag
\end{align}

If $S(\lambda)=I+s\sigma_{\pm}$ then
$G(\lambda)=I+se^{\pm2g}\sigma_{\pm}$.  Our next goal is to transform
the jump contour $\gamma$ to the contour of the steepest descent for
the matrix $G(\lambda)-I$. We denote by $\gamma_+$ the level line $\Im
g(\lambda)=const$ passing through the stationary phase point
$\lambda=\lambda_0=\sqrt{e^{-i\pi}x/6}$ and asymptotic to the rays
$\arg\lambda=\pm\frac{2\pi}{5}$. This is the steepest descent line for
$e^{2g}$. Let $\gamma_-=\cup_j\ell_j\cup\sigma$ be the union of the
level lines $\ell_j$, $j=0,1,2$, $\Im g(\lambda)=const$, and $\sigma$,
$\Re g(\lambda)=const$, all emanating from the critical point
$\lambda=-2\lambda_0$. Among them, the level line $\ell_1$ approaching
the ray $\arg\lambda=\frac{2\pi}{5}$ (if $\arg x=\pi$, the level line
$\ell_1$ is the segment $[-2\lambda_0,\lambda_0]$) is the steepest
descent line for $e^{2g}$, while the level lines $\ell_0$ and $\ell_2$
approaching the rays $\arg\lambda=-\frac{4\pi}{5}$ and
$\arg\lambda=\frac{4\pi}{5}$, respectively, are the steepest descent
lines for $e^{-2g}$. The level line $\sigma$, $\Re g(\lambda)=const$,
approaches the ray $\arg\lambda=\pi$.

\begin{figure}[hbt]
\begin{center}
\epsfig{file=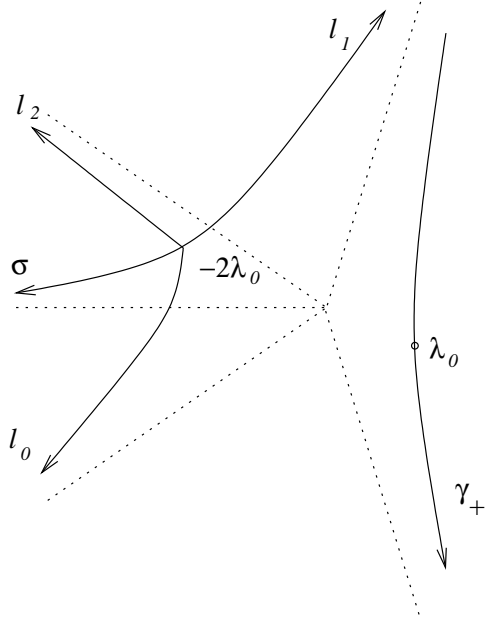}
\end{center}
\caption{A RH problem graph for $s_0=0$.}
\label{f2}
\end{figure}

Since the Stokes matrix $S_1$ can be factorized,
$$
S_1=\begin{pmatrix}
1&s_1\\
0&1
\end{pmatrix}=
\begin{pmatrix}
1&i-s_{-1}\\
0&1
\end{pmatrix}=S_{-1}^{-1}
\begin{pmatrix}
1&i\\
0&1
\end{pmatrix},
$$
it is convenient to consider the following equivalent RH problems for
$\Psi(\lambda)$:

\smallskip
\noindent
for $\arg x\in[\frac{3\pi}{5},\pi]$, the jump contour is the union of
$\gamma_+$ oriented from up to down and $\gamma_-$ whose components
are oriented toward infinity, see Figure~\ref{f2}. The jump matrices 
are as follows:
\begin{alignat}{2}\label{+jumps}
&\lambda\in\gamma_+\colon\quad&& S(\lambda)=S_{-1}=
\begin{pmatrix}
1&s_{-1}\\ 0&1
\end{pmatrix},\notag
\\ &\lambda\in\ell_1\colon\quad&& S(\lambda)=S_+=
\begin{pmatrix}
1&i\\ 0&1
\end{pmatrix},\notag
\\ &\lambda\in\ell_0\cup\ell_2\colon\quad&& S(\lambda)=S_-=
\begin{pmatrix}
1&0\\ i&1
\end{pmatrix},\notag
\\ &\lambda\in\sigma\colon\quad&& S(\lambda)=-i\sigma_1.
\end{alignat}

\begin{remark}\label{rem2}
The jump contour for the RH problem (\ref{+jumps}) is decomposed into
the disjoint union of the line $\gamma_+$ and the graph $\gamma_-$, 
see Figure~\ref{f2}. For the boundary value $\arg x=\pi-0$, the level line 
$\ell_1$ emanating from $\lambda=-2\lambda_0$ passes through the 
stationary phase point $\lambda=\lambda_0$ and partially merges with 
the upper half of the level line $\gamma_+$. To construct the RH 
problem however, it is not necessary to transform the jump contour 
precisely to the steepest descent graph. It is enough to ensure that 
the jump matrices approach the unit matrix uniformly with respect to 
$\lambda$ and fast enough as $x\to\infty$ or $\lambda\to\infty$.
\end{remark}

As $\arg x\in[\frac{3\pi}{5},\pi]$, introduce the reduced RH problem
($s_0=s_{-1}=0$) for the piece-wise holomorphic function
$\Phi(\lambda)$ discontinuous across $\gamma_-$ only:
\begin{align}\label{Phi_RH}
i)\quad&
\lim_{\lambda\to\infty}\lambda^{1/2}\Bigl(
\tfrac{1}{\sqrt2}(\sigma_3+\sigma_1)\lambda^{-\frac{1}{4}\sigma_3}
\Phi(\lambda)e^{-\theta\sigma_3}-I\bigr)
\quad\text{is diagonal},\notag 
\\
ii)\quad&\Phi_+(\lambda)=\Phi_-(\lambda)S(\lambda),\quad
\lambda\in\gamma_-=\cup_{j=0,1,2}\ell_j\cup\sigma.
\end{align}
The jump matrix $S(\lambda)$ here is defined in (\ref{+jumps}).

\begin{theorem}\label{Theorem1}
If\/ $\arg x\in[\frac{3\pi}{5},\pi]$ and $|x|$ is large enough, then
there exists a unique solution of the RH problem (\ref{Phi_RH}). The
Painlev\'e function $y_0(x)$ corresponding to $s_0=s_{-1}=0$ has the 
asymptotics $y_0(x)=\sqrt{e^{-i\pi}x/6}+{\cal O}(x^{-2})$ as 
$x\to\infty$ in the above sector.
\end{theorem}
\proof Uniqueness.  Since $\det S(\lambda)\equiv1$, we have
$\det\Phi_+=\det\Phi_-$, and hence $\det\Phi(\lambda)$ is an entire
function. Furthermore, because of normalization of $\Phi(\lambda)$ at
infinity, $\det\Phi(\lambda)\equiv-1$.  Let $\tilde\Phi$ and $\Phi$ be
two solutions of (\ref{Phi_RH}). Taking into account the cyclic
relation in (\ref{Stokes_relations}) which implies the continuity of
the RH problem for $\Phi(\lambda)$ at $\lambda=-2\lambda_0$, the
``ratio" $\chi(\lambda)=\tilde\Phi(\lambda)\Phi^{-1}(\lambda)$ is an
entire function of $\lambda$. Using the Liouville theorem and
normalization of $\Phi$ and $\tilde\Phi$ at infinity, we find
$\chi(\lambda)\equiv I$,
i.e. $\Phi(\lambda)\equiv\tilde\Phi(\lambda)$.

Existence. Introduce an auxiliary function
\begin{equation}\label{hat_Phi_0r}
\hat\Phi_0(z)=
\begin{pmatrix}
v_1'(z)&v_2'(z)\\ v_1(z)&v_2(z)
\end{pmatrix},
\end{equation}
where the prime means differentiation w.r.t.\ $z$ and
\begin{equation}\label{y12}
v_1(z)=\sqrt{2\pi}\,e^{i\pi/6}\Ai\bigl(e^{i2\pi/3}z\bigr),\quad
v_2(z)=-\sqrt{2\pi}\,\Ai(z),
\end{equation}
with $\Ai(z)$ standing for the classical Airy function which can be
defined using the Taylor expansion \cite{BE, olver},
\begin{equation}\label{Ai}
\Ai(z)=\frac{1}{3^{2/3}\Gamma(\frac{2}{3})} 
\sum_{k=0}^{\infty}
\frac{3^k\Gamma(k+\frac{1}{3})z^{3k}}{\Gamma(\frac{1}{3})(3k)!}
-\frac{1}{3^{1/3}\Gamma(\frac{1}{3})} 
\sum_{k=0}^{\infty}
\frac{3^k\Gamma(k+\frac{2}{3})z^{3k+1}}{\Gamma(\frac{2}{3})(3k+1)!}.
\end{equation}
Asymptotics at infinity of this function and its derivative are as
follows,
\begin{multline}\label{Ai_as}
\Ai(z)=\tfrac{1}{2\sqrt\pi}z^{-1/4}e^{-\frac{2}{3}z^{3/2}}
\Bigl\{\sum_{n=0}^N(-1)^n3^{-2n}
\frac{\Gamma(3n+\frac{1}{2})}{\Gamma(\frac{1}{2})(2n)!}z^{-3n/2}
+{\cal O}\bigl(z^{-3(N+1)/2}\bigr)\Bigr\}, 
\\
\Ai'(z)=\tfrac{1}{2\sqrt\pi}z^{1/4}e^{-\frac{2}{3}z^{3/2}} 
\Bigl\{\sum_{n=0}^N(-1)^n3^{-2n}\bigl(3n+\tfrac{1}{2}\bigr)
\frac{\Gamma(3n-\frac{1}{2})}{\Gamma(\frac{1}{2})(2n)!}z^{-\frac{3n}{2}}
+{\cal O}\bigl(z^{-3(N+1)/2}\bigr) 
\Bigr\}, 
\\ 
\text{as}\quad
z\to\infty,\quad 
\arg z\in(-\pi,\pi).
\end{multline}
It is worth to note that the function $\hat\Phi_0(z)$ satisfies the
linear differential equation
\begin{equation}\label{Phi0_eq}
\frac{d\hat\Phi_0}{dz}= \bigl\{z\sigma_++\sigma_-\bigr\}\hat\Phi_0.
\end{equation}
Using the properties of the Airy functions, we find that the products
\begin{equation}\label{hat_Phi_k}
\hat\Phi_1(z)=\hat\Phi_0(z)S_-,\quad
\hat\Phi_2(z)=\hat\Phi_1(z)S_+,\quad \hat\Phi_3(z)=\hat\Phi_2(z)S_-,
\end{equation}
$S_{\pm}=I+i\sigma_{\pm}$, have the asymptotic expansion
\begin{equation}\label{hat_Phi_k_as}
\hat\Phi_k(z)=
z^{\frac{1}{4}\sigma_3}\tfrac{1}{\sqrt2}(\sigma_3+\sigma_1)
V_{\infty}(z)e^{\frac{2}{3}z^{3/2}\sigma_3},
\end{equation}
as $|z|\to\infty$, $\arg z\in
\bigl(-\pi+\tfrac{2\pi}{3}k,\tfrac{\pi}{3}+\tfrac{2\pi}{3}k)\bigr)$,
where
\begin{equation}\label{V_8}
V_{\infty}(z)=I-\sum_{n=1}^{\infty}
3^{-2n}\frac{\Gamma(3n-\frac{1}{2})}{2\Gamma(\frac{1}{2})(2n)!}
z^{-3n/2}
\begin{pmatrix}
1&(-1)^n6n\\ 6n&(-1)^n
\end{pmatrix}.
\end{equation}

Let $\hat\gamma_-=\hat\sigma\cup_{j=0,1,2}\hat\ell_j$ be the union 
of the rays
$\hat\ell_j=\{z\in{\Bbb C}\colon\arg z=\frac{2\pi}{3}(j-1)\}$,
$j=0,1,2$, and $\hat\sigma=\{z\in{\Bbb C}\colon\arg z=\pi\}$ all
oriented toward infinity. This graph divides the complex $z$-plane
into four regions: $\hat\omega_0$ which is the sector between
$\hat\sigma$ and $\hat\ell_0$, the sectors $\hat\omega_k$, $k=1,2$,
between the rays $\hat\ell_{k-1}$ and $\hat\ell_k$, and the sector
$\hat\omega_3$ between the rays $\hat\ell_2$ and $\hat\sigma$. 

\begin{figure}[hbt]\begin{center}
\epsfig{file=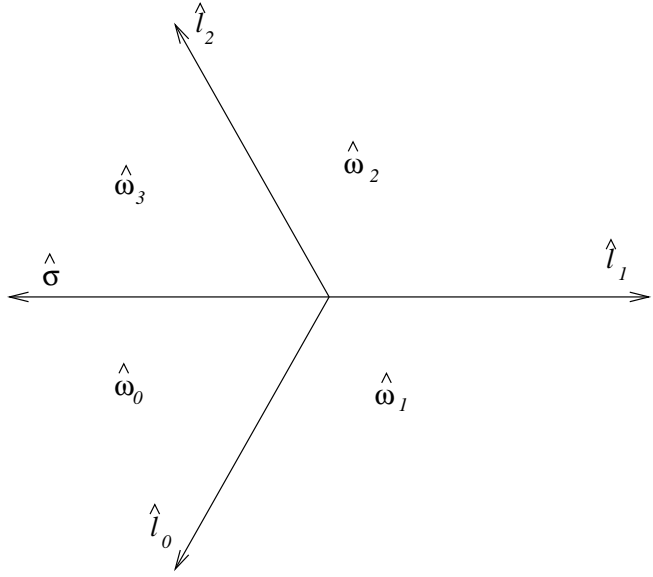}
\end{center}
\caption{The model RH problem graph.}
\label{f3}
\end{figure}

Define a piece-wise holomorphic function $\hat\Phi(z)$,
\begin{equation}\label{hat_Phi_collect}
\hat\Phi(z)\Bigr|_{z\in\hat\omega_k}= \hat\Phi_k(z).
\end{equation}
By construction, this function solves the following RH problem,
see Figure~\ref{f3}:
\begin{align}\label{hat_Phi_as_RH}
i)\quad& 
\tfrac{1}{\sqrt2}(\sigma_3+\sigma_1) z^{-\frac{1}{4}\sigma_3}
\hat\Phi(z) e^{-\frac{2}{3}z^{3/2}\sigma_3}=I+{\cal O}(z^{-3/2}),\quad 
z\to\infty, 
\\
ii)\quad& 
z\in\hat\gamma_-\colon\quad 
\hat\Phi_+(z)=\hat\Phi_-(z)\hat S(z),\notag 
\\\label{hat_jumps} 
&z\in\hat\ell_1\colon\quad 
\hat S(z)=S_+,\qquad 
z\in\hat\ell_0\cup\hat\ell_2\colon\quad 
\hat S(z)=S_-,\notag 
\\ 
&z\in\hat\sigma\colon\quad 
\hat S(z)=-i\sigma_1.
\end{align}

Therefore the function $\hat\Phi(z)$ has precisely the jump properties
of the function $\Phi(\lambda)$. To find $\Phi(\lambda)$ with the
correct asymptotic behavior at infinity, let us use the mapping
\begin{multline}\label{z_lambda}
\tfrac{2}{3}z^{3/2}=g(\lambda)= \tfrac{4}{5}(\lambda+2\lambda_0)^{5/2}
-4\lambda_0(\lambda+2\lambda_0)^{3/2},\quad\text{or}\quad \\
z(\lambda)=(-6\lambda_0)^{2/3}(\lambda+2\lambda_0)
\bigl(1-\tfrac{1}{5\lambda_0}(\lambda+2\lambda_0)\bigr)^{2/3},\quad
\lambda_0=\sqrt{e^{-i\pi}x/6},
\end{multline}
Within the disk $|\lambda+2\lambda_0|\leq
R<3|\lambda_0|=|\frac{3}{2}x|^{1/2}$, the mapping (\ref{z_lambda})
yields a holomorphic change of the independent variable. Introduce a
piece-wise holomorphic function $\tilde\Phi(\lambda)$,
\begin{align}\label{tilde_Phi}
\tilde\Phi(\lambda)= &\begin{cases} B(\lambda)
\hat\Phi(z(\lambda)),\quad |\lambda+2\lambda_0|<R,\\
(\lambda+2\lambda_0)^{\frac{1}{4}\sigma_3}
\frac{1}{\sqrt2}(\sigma_3+\sigma_1) e^{g(\lambda)\sigma_3},\quad
|\lambda+2\lambda_0|>R,
\end{cases}
\\ &B(\lambda)= (-6\lambda_0)^{-\frac{1}{6}\sigma_3}
\bigl(1-\tfrac{\lambda+2\lambda_0}{5\lambda_0}\bigr)^{-\frac{1}{6}\sigma_3},
\notag
\end{align}
where $(\lambda+2\lambda_0)^{1/4}$ is defined on the plane cut along
the level line $\sigma$ asymptotic to the ray $\arg\lambda=\pi$. Note
that $B(\lambda)$ is holomorphic in the interior of the above disk
$|\lambda+2\lambda_0|\leq R<3|\lambda_0|$ and thus does not affect the
jump properties of $\hat\Phi(z(\lambda))$. We look for the solution of
the RH problem (\ref{Phi_RH}) in the form of the product
\begin{equation}\label{chi_def}
\Phi(\lambda)=(I+(4\lambda_0^3-{\Eu H})\sigma_+)
\chi(\lambda)\tilde\Phi(\lambda).
\end{equation}

\begin{figure}[hbt]\begin{center}
\epsfig{file=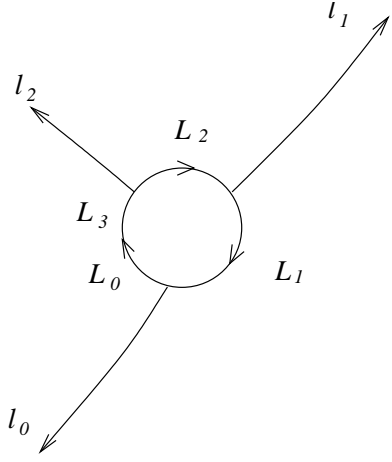}
\end{center}
\caption{A RH problem graph for the correction 
function $\chi(\lambda)$.}
\label{f4}
\end{figure}

Consider the RH problem for the correction function $\chi(\lambda)$. By
construction, it is a piece-wise holomorphic function discontinuous 
across the clockwise oriented circle ${\cal L}$ of the radius $R$ 
centered at $-2\lambda_0$ and across the part of $\gamma_-$ located 
outside the above circle (in fact, $\chi(\lambda)$ is continuous across
$\sigma$, see (\ref{chi_RH}) below). The latter is divided by $\gamma_-$ 
in four arcs: ${\cal L}_0$ between $\sigma$ and $\ell_0$, ${\cal L}_k$, 
$k=1,2$, between $\ell_{k-1}$ and $\ell_k$, and ${\cal L}_3$ between 
$\ell_2$ and $\sigma$, see Figure~\ref{f4}. To simplify our notations, 
let us put
\begin{equation}\label{tilde_lambda}
\tilde\lambda=\lambda+2\lambda_0.
\end{equation}
Then the RH problem for $\chi(\lambda)$ is as follows:
\begin{align}\label{chi_RH}
i)\quad 
&\chi(\lambda)\to I,\quad &&\lambda\to\infty; \notag \\
ii)\quad 
&\chi^+(\lambda)=\chi^-(\lambda){\cal G}(\lambda),
&&\lambda\in\ell,\quad\text{where} \notag \\ 
&\lambda\in\ell_1,\
|\tilde\lambda|>R\colon 
&&{\cal G}(\lambda)=I+\tfrac{i}{2}e^{2g}
\bigl(\sigma_3-\tilde\lambda^{1/2}\sigma_+
+\tilde\lambda^{-1/2}\sigma_-\bigr),\notag \\
&\lambda\in\ell_0\cup\ell_2,\ 
|\tilde\lambda|>R\colon 
&&{\cal G}(\lambda)=I+\tfrac{i}{2}e^{-2g}\bigl(
\sigma_3+\tilde\lambda^{1/2}\sigma_+
-\tilde\lambda^{-1/2}\sigma_-\bigr),\notag \\ 
&\lambda\in\sigma,\ 
|\tilde\lambda|>R\colon 
&&{\cal G}(\lambda)=I, \notag \\ 
&|\tilde\lambda|=R,\ 
\lambda\in{\cal L}_k\colon\ 
&&{\cal G}(\lambda)=
B(\lambda)\hat\Phi_k(z(\lambda))e^{-g\sigma_3}
\tfrac{1}{\sqrt2}(\sigma_3+\sigma_1)
\tilde\lambda^{-\frac{1}{4}\sigma_3}, \notag \\ 
&&& k=0,1,2,3.
\end{align}

Taking into account the equations (\ref{hat_jumps}), it is easy to
check the continuity of the RH problem at the node points. Observing
that, on the circle $|\tilde\lambda|=R=c|x|^{1/2}$, $0<c<\sqrt{3/2}$,
we have $z(\lambda)={\cal O}(|x|^{5/6})$ is large, we immediately see
that
\begin{equation}\label{Hk_estim}
\|{\cal G}(\lambda)-I\|\leq
c|\tilde\lambda|^{1/2}e^{-(2/3)^{1/2}|x|^{1/2}\,|\tilde\lambda|^{3/2}},\quad
\lambda\in\ell_k,\quad k=0,1,2,\quad |\tilde\lambda|\geq R,
\end{equation}
where the precise value of the positive constant $c$ is not important
for us. Taking into account that, by the above reason, on the circle
$|\tilde\lambda|=R$, we may use for $\hat\Phi_k$ its asymptotics
(\ref{hat_Phi_k_as}), the jump matrix ${\cal G}(\lambda)$ has the
asymptotic expansion
\begin{multline}\label{H_on_circle}
{\cal G}(\lambda)-I= \\ =-\sum_{n=1}^{\infty}
3^{-2n}\frac{\Gamma(3n-\frac{1}{2})}{2\Gamma(\frac{1}{2})(2n)!}
z^{-\frac{3n}{2}}\begin{pmatrix} \frac{1+(-1)^n}{2}(1+6n)&
\frac{1-(-1)^n}{2}(1+6n)\tilde\lambda^{1/2} \\
\frac{1-(-1)^n}{2}(1-6n)\tilde\lambda^{-1/2}& \frac{1+(-1)^n}{2}(1-6n)
\end{pmatrix},
\\
z=(-6\lambda_0)^{2/3}\bigl(1-\frac{\tilde\lambda}{5\lambda_0}\bigr)^{2/3}
\tilde\lambda,\quad \tilde\lambda=\lambda+2\lambda_0,\quad
|\tilde\lambda|=R.
\end{multline}
Therefore, we have the estimate
\begin{equation}\label{H_on_circle_estim}
\|{\cal G}(\lambda)-I\|\leq cR^{-2}=c'|x|^{-1},
\end{equation}
where the precise value of the positive constants $c,c'$ is not
important for us.

Now, the solubility of the RH problem (\ref{chi_RH}) and therefore of
(\ref{Phi_RH}) for large enough $|x|$ is straightforward. Indeed,
consider the equivalent system of the non-homogeneous singular
integral equations for the limiting value $\chi^+(\lambda)$, i.e.\
\begin{equation}\label{chi_sing}
\chi^+(\lambda)= I-\frac{1}{2\pi i}\int_{\ell}
\frac{\chi^+(\zeta)({\cal G}^{-1}(\zeta)-I)}{\zeta-\lambda_+}\,d\zeta,
\end{equation}
or, in the symbolic form, $\chi^+=I+K\chi^+$. Here $\lambda_+$ means
the left limit of $\lambda$ on $\ell$ (recall, that the circle
$|\tilde\lambda|=R$ is clock-wise oriented), and $K$ is the
composition of the operator of the right multiplication in ${\cal
G}^{-1}(\lambda)-I$ and of the Cauchy operator $C_+$. An equivalent
singular integral equation for $\psi^+:=\chi^+-I$ differs from
(\ref{chi_sing}) in the inhomogeneous term only,
\begin{equation}\label{psi_sing}
\psi^+=KI+K\psi^+.
\end{equation}
Consider the integral equation (\ref{psi_sing}) in the space
$L_2(\ell)$.  Since ${\cal G}^{-1}(\lambda)-I$ is small in $L_2(\ell)$
for large enough $|x|$, and $C_+$ is bounded in $L_2(\ell)$, then
$\|K\|_{L_2(\ell)}\leq c|x|^{-1/2}$ with some positive constant $c$,
thus $K$ is contracting and $I-K$ is invertible in $L_2(\ell)$ for large
enough $|x|$. Because $KI\in L_2(\ell)$, equation (\ref{psi_sing}) for
$\psi^+$ is solvable in $L_2(\ell)$, and the solution $\chi(\lambda)$
of the RH problem (\ref{chi_RH}) is determined by $\psi^+(\lambda)$
using the equation $\chi^+=I+KI+K\psi^+$.

Let us find the asymptotics of the Painlev\'e function.  Using
(\ref{Y_expansion}) and definition of $\tilde\Phi(\lambda)$
(\ref{tilde_Phi}), the asymptotics of $\chi(\lambda)$ as
$\lambda\to\infty$ in terms of $y$ and ${\Eu H}$ is as follows,
\begin{multline}\label{chi_expansion}
\chi(\lambda)=I
+\frac{1}{2\lambda}\bigl(y-\lambda_0-(4\lambda_0^3-{\Eu H})^2\bigr)
\sigma_3 +\frac{1}{\lambda}(4\lambda_0^3-{\Eu H})\sigma_- + \\
+\begin{pmatrix} {\cal O}(\lambda^{-3/2})& {\cal O}(\lambda^{-1})\\
{\cal O}(\lambda^{-2})& {\cal O}(\lambda^{-3/2})
\end{pmatrix}.
\end{multline}

On the other hand, in accord with the said above, the function
$\chi^+$ is given by the converging iterative series,
$\chi^+=\sum_{n=0}^{\infty}K^nI$. To compute the term $K^nI$, we
observe that the contribution of the infinite branches $\ell_k$ is
exponentially small in $x$ due to estimate (\ref{Hk_estim}). Using the
expansion (\ref{H_on_circle}), we reduce the evaluation of the
integral along the circle $|\tilde\lambda|=R$ to the residue
theorem. Omitting this elementary computation, we present the final
result: For large enough $|x|$, $\arg x\in[\frac{3\pi}{5},\pi]$, the
asymptotics of $\chi(\lambda)$ as $\lambda\to\infty$ is given by
\begin{equation}\label{chi_as}
\chi(\lambda)=I 
+\begin{pmatrix} 
{\cal O}(\lambda_0^{-4}\tilde\lambda^{-1})& 
{\cal O}(\lambda_0^{-1}\tilde\lambda^{-1})\\ 
{\cal O}(\lambda_0^{-2}\tilde\lambda^{-1})& 
{\cal O}(\lambda_0^{-4}\tilde\lambda^{-1})
\end{pmatrix}.
\end{equation}

Comparing entries $\chi_{21}(\lambda)$ in (\ref{chi_expansion}) and
(\ref{chi_as}), we see that the Hamiltonian function ${\Eu H}={\Eu H}_0(x)$ 
corresponding to the Stokes multipliers $s_0=s_{-1}=0$ is given by
\begin{equation}\label{Ham0_as}
{\Eu H}={\Eu H}_0(x)=4\lambda_0^3+{\cal O}(\lambda_0^{-2})=
4(-x/6)^{3/2}+{\cal O}(x^{-1}).
\end{equation}
Next comparing entries $\chi_{11}(\lambda)$ in (\ref{chi_expansion})
and (\ref{chi_as}) and using (\ref{Ham0_as}), we find the asymptotics 
of the Painlev\'e function $y=y_0(x)$,
\begin{equation}\label{y0_as}
y=y_0(x)=\lambda_0+{\cal O}(\lambda_0^{-4})= \sqrt{-x/6}+{\cal
O}(x^{-2}).
\end{equation}
Recall that $\lambda_0=(e^{-i\pi}x/6)^{1/2}$ where the main branch of
the root is taken.  \endproof

Let us go to the case of the nontrivial $s_{-1}$ described by the RH
problem (\ref{+jumps}). We look for the solution $\Psi(\lambda)$ in
the form of the product
\begin{equation}\label{Psi_chi}
\Psi(\lambda)= \bigl(I-({\Eu H}-{\Eu H}_0)\sigma_+\bigr)
X(\lambda)\Phi(\lambda),
\end{equation}
where $\Phi(\lambda)$ is the solution of the reduced RH problem
(\ref{Phi_RH}) and ${\Eu H}_0$ (\ref{Ham0_as}) is the Hamiltonian
function (\ref{Ham_def}) corresponding to the Painlev\'e transcendent
$y_0(x)$ (\ref{y0_as}). Using (\ref{Y_expansion}), we find the
asymptotics of $X(\lambda)$ as $\lambda\to\infty$,
\begin{multline}\label{X_expansion}
X(\lambda)=\bigl(I+({\Eu H}-{\Eu H}_0)\sigma_+\bigr)\Psi\Phi^{-1}= \\
=I +\frac{1}{2\lambda}\bigl(y-y_0-({\Eu H}-{\Eu H}_0)^2)\sigma_3
-\frac{1}{\lambda}({\Eu H}-{\Eu H}_0)\sigma_- +\begin{pmatrix} {\cal
O}(\lambda^{-3/2})& {\cal O}(\lambda^{-1})\\ {\cal O}(\lambda^{-2})&
{\cal O}(\lambda^{-3/2})
\end{pmatrix}.
\end{multline}

Thus we arrive at the RH problem for the correction function
$X(\lambda)$ on the steepest descent line $\gamma_+$,
\begin{align}\label{Psi_X_RH}
i)&\quad X(\lambda)\to I,&\quad \lambda\to\infty,&&\notag \\ ii)&\quad
X_+(\lambda)=X_-(\lambda){\Eu G}(\lambda),&\quad \lambda\in\gamma_+,&&
\\ &\qquad {\Eu G}(\lambda)=
\Phi(\lambda)S_{-1}\Phi^{-1}(\lambda).&&&\notag
\end{align}
Note, $\Phi(\lambda)$ is continuous across $\gamma_+$ and therefore
holomorphic in some neighborhood of $\gamma_+$ as $\arg
x\in[\frac{3\pi}{5},\pi]$, $|x|$ is large enough. 

The jump matrix on $\gamma_+$ can be estimated as follows,
\begin{equation}\label{X_H_estim}
\|{\Eu G}(\lambda)-I\|\leq c|s_{-1}|
e^{-\frac{1}{5}2^{11/4}3^{1/4}|x|^{5/4}\cos(\frac{5}{4}(\arg x-\pi))}
e^{-2^{3/4}3^{1/4}|x|^{1/4}|\lambda-\lambda_0|^2}.
\end{equation}
Here $c$ is some positive constant whose precise value is not
important for us, and $\lambda_0=(e^{-i\pi}x)^{1/2}$ is the stationary
phase point for $\exp(g(\lambda))$, see (\ref{g_function}). Estimate
(\ref{X_H_estim}) yields the estimate for the norm of the singular
integral operator ${\cal K}$ in the equivalent system of singular
integral equations, $X_-=I+{\cal K}X_-$,
\begin{equation}\label{K_estim}
\|{\cal K}\|_{L_2(\gamma_+)}\leq
c'|s_{-1}|e^{-\frac{1}{5}2^{11/4}3^{1/4}|x|^{5/4}
\cos(\frac{5}{4}(\arg x-\pi))},\quad c'>0.
\end{equation}
If $|x|$ is large enough and $\arg x\in[\frac{3\pi}{5}+\epsilon,\pi]$,
$\epsilon>0$, then the operator ${\cal K}$ is contracting and the
system $X_-=I+{\cal K}X_-$ is solvable by iterations in
$L_2(\gamma_+)$, i.e.\ $X_-=\sum_{n=0}^{\infty}{\cal
K}^nX_-$. However, to incorporate the oscillating direction $\arg
x=\frac{3\pi}{5}$ in the general scheme, we use some more refined
procedure.

\begin{theorem}\label{Theorem2}
If $s_0=0$, $\arg x\in[\frac{3\pi}{5},\pi]$ and $|x|$ is large enough, then
there exists a unique solution of the RH problem
(\ref{p28})--(\ref{jump_matrices}). The 
corresponding Painlev\'e function has the asymptotics 
\begin{equation}\label{y+_as}
y(x)=y_0(x) +\frac{s_{-1}}{\sqrt\pi}2^{-11/8}3^{-1/8}
(e^{-i\pi}x)^{-1/8}e^{-\frac{1}{5}2^{11/4}3^{1/4}(e^{-i\pi}x)^{5/4}}
(1+{\cal O}(x^{-3/8})),
\end{equation}
where $y_0(x)\sim\sqrt{e^{-i\pi}x/6}$ is the solution of the
Painlev\'e equation for $s_0=s_{-1}=0$, $s_1=s_2=s_{-2}=i$.
\end{theorem}

\proof It is enough to prove the solubility of the RH problem 
(\ref{Psi_X_RH}).

Using for $\Phi(\lambda)$ the expressions (\ref{chi_def}) with
(\ref{tilde_Phi}) and the estimate (\ref{chi_as}) together, we find
the asymptotics of the jump matrix ${\Eu G}(\lambda)$,
\begin{multline}\label{H_as}
{\Eu G}(\lambda)= I+\tfrac{1}{2}s_{-1}e^{2g}
\begin{pmatrix}
1+ {\cal O}(\lambda_0^{-2}\tilde\lambda^{-1/2})& -\tilde\lambda^{1/2}
+{\cal O}(\lambda_0^{-2})\\ \tilde\lambda^{-1/2} +{\cal
O}(\lambda_0^{-2}\tilde\lambda^{-1})& -1+{\cal
O}(\lambda_0^{-2}\tilde\lambda^{-1/2})
\end{pmatrix},
\\ \lambda\in\gamma_+,\quad \tilde\lambda=\lambda+2\lambda_0.
\end{multline}

Consider the following model RH problem,
\begin{align}\label{P_RH}
i)\quad &P(\lambda)\to I,\quad \lambda\to\infty,\notag \\ ii)\quad
&P_+(\lambda)=P_-(\lambda)\hat{\Eu G}(\lambda),\quad
\lambda\in\gamma_+, \\ &\hat{\Eu
G}(\lambda)=I+\tfrac{1}{2}s_{-1}e^{2g}
\begin{pmatrix}
1&-(3\lambda_0)^{1/2}\\ (3\lambda_0)^{-1/2}&-1
\end{pmatrix}.\notag
\end{align}
This problem is solvable by the following quadrature,
\begin{equation}\label{P_explicit}
P(\lambda)=I+\tfrac{1}{2}s_{-1} \tfrac{1}{2\pi
i}\int_{\gamma_+}\frac{e^{2g}}{\zeta-\lambda}\,d\zeta
\begin{pmatrix}
1&-(3\lambda_0)^{1/2}\\ (3\lambda_0)^{-1/2}&-1
\end{pmatrix}.
\end{equation}

We look for the solution $X(\lambda)$ of the RH problem (\ref{Psi_X_RH})
in the form of the product,
\begin{equation}\label{X=QP}
X(\lambda)=Q(\lambda)P(\lambda).
\end{equation}
The correction function $Q(\lambda)$ satisfies the RH problem
\begin{align}\label{Q_RH}
i)&\quad Q(\lambda)\to I,&\quad \lambda\to\infty,&&\notag \\ ii)&\quad
Q_+(\lambda)=Q_-(\lambda)W(\lambda),&\quad \lambda\in\gamma_+,&&\notag
\\ &\qquad W(\lambda)=P_-(\lambda){\Eu G}(\lambda) \hat{\Eu
G}(\lambda)^{-1}P_-^{-1}(\lambda).&&&
\end{align}

Using (\ref{H_as})--(\ref{P_explicit}), we find the estimate for the
jump matrix $W(\lambda)$ on $\gamma_+$,
\begin{equation}\label{W_as}
W(\lambda)=I+{\cal
O}(s_{-1}e^{2g}(\lambda-\lambda_0)\lambda_0^{-1/2}),\quad
\lambda\in\gamma_+,
\end{equation}

Our next steps are similar to presented in the proof of
Theorem~\ref{Theorem1}. Consider the system of the singular integral
equations for $Q_+(\lambda)$ equivalent to the RH problem
(\ref{Q_RH}), $Q_+=I+{\Eu K}Q_+$. Here the singular integral operator
${\Eu K}$ is the superposition of the multiplication operator in $W-I$
and of the Cauchy operator $C_+$. Because the Cauchy operator is
bounded in $L_2(\gamma_+)$, the singular integral operator ${\Eu K}$
for large enough $|x|$, $\arg x\in[\frac{3\pi}{5},\pi]$, satisfies the
estimate
\begin{equation}\label{Eu_K_estim}
\|{\Eu K}\|_{L_2(\gamma_+)}\leq
c|s_{-1}||x|^{-1/2}e^{-\frac{1}{5}2^{11/4}3^{1/4}|x|^{5/4}
\cos(\frac{5}{4}(\arg x-\pi))},
\end{equation}
with some positive constant $c$ whose precise value is not important
for us. Thus equation $\zeta_+={\Eu K}I+{\Eu K}\zeta_+$ for the
difference $\zeta_+:=Q_+-I$ is solvable by iterations in the space
$L_2(\gamma_+)$ for large enough $|x|$. Solution of the RH problem
(\ref{Q_RH}) is given by the integral $Q=I+{\Eu K}I+{\Eu
K}\zeta_+$. This implies the asymptotics of $Q(\lambda)$ as
$\lambda\to\infty$,
\begin{multline}\label{Q_as}
Q(\lambda)=I+\tfrac{1}{2\pi i}\int_{\gamma_+} \bigl(I+{\cal O}({\Eu
K}I(\zeta))\bigr)(I-W^{-1}(\zeta)) \frac{d\zeta}{\zeta-\lambda}= \\
=I+{\cal O}\bigl(\lambda^{-1}s_{-1}x^{-1/2}
\exp(-\tfrac{1}{5}2^{11/4}3^{1/4}\Re(e^{-i\pi}x)^{5/4})\bigr).
\end{multline}

Now let us find the asymptotics of the Painlev\'e function
$y(x)$. Using (\ref{X=QP}), (\ref{P_explicit}) and the estimate
(\ref{Q_as}), we find
\begin{multline}\label{X_as}
X(\lambda)=I+\frac{s_{-1}}{\lambda\sqrt\pi}
2^{-19/8}3^{-1/8}(e^{-i\pi}x)^{-1/8}
e^{-\frac{1}{5}2^{11/4}3^{1/4}(e^{-i\pi}x)^{5/4}} \bigl(I+{\cal
O}(x^{-3/8})\bigr)\times \\ \times
\begin{pmatrix}
1&-2^{-1/4}3^{1/4}(e^{-i\pi}x)^{1/4}\\
2^{1/4}3^{-1/4}(e^{-i\pi}x)^{-1/4}&-1
\end{pmatrix}.
\end{multline}
Comparing (\ref{X_as}) and (\ref{X_expansion}), we conclude that
the Hamiltonian function for $s_0=0$ is as follows,
\begin{equation}\label{Ham_as}
{\Eu H}(x)={\Eu H}_0(x) -\frac{s_{-1}}{\sqrt\pi}
2^{-17/8}3^{-3/8}(e^{-i\pi}x)^{-3/8}
e^{-\frac{1}{5}2^{11/4}3^{1/4}(e^{-i\pi}x)^{5/4}} \bigl(1+{\cal
O}(x^{-1/8})\bigr),
\end{equation}
while the Painlev\'e function is given by
\begin{equation}\label{y_as}
y(x)=y_0(x) +\frac{s_{-1}}{\sqrt\pi}
2^{-11/8}3^{-1/8}(e^{-i\pi}x)^{-1/8}
e^{-\frac{1}{5}2^{11/4}3^{1/4}(e^{-i\pi}x)^{5/4}} \bigl(1+{\cal
O}(x^{-3/8})\bigr),
\end{equation}
where ${\Eu H}_0(x)$ and $y_0(x)$ are the Hamiltonian and the
Painlev\'e functions, respectively, corresponding to $s_0=s_{-1}=0$.
\endproof

\subsection{Other degenerate Painlev\'e functions}

Applying the symmetry (\ref{P_symmetries_bar}) to the solution 
(\ref{y+_as}) and changing the argument of $x$ in $2\pi$, we obtain
\begin{theorem}\label{Theorem3}
If $s_0=0$ and $|x|\to\infty$, $\arg x\in[\pi,\frac{7\pi}{5}]$, then
the asymptotics of the Painlev\'e first transcendent is given by
\begin{equation}\label{y-_as}
y(x)=y_1(x) -\frac{s_{1}}{\sqrt\pi}2^{-11/8}3^{-1/8}
(e^{-i\pi}x)^{-1/8}e^{-\frac{1}{5}2^{11/4}3^{1/4}(e^{-i\pi}x)^{5/4}}
(1+{\cal O}(x^{-3/8})),
\end{equation}
where $y_1(x)\sim\sqrt{e^{-i\pi}x/6}$ is the solution of the
Painlev\'e equation for $s_0=s_{1}=0$, $s_{-1}=s_2=s_{-2}=i$.
\end{theorem}

The solutions $y_0(x)$ and $y_1(x)=\overline{y_0(e^{2\pi i}\bar x)}$
are meromorphic functions of $x\in{\Bbb C}$ and thus can be continued
beyond the sectors indicated in Theorems~\ref{Theorem2}
and~\ref{Theorem3}. To find the asymptotics of $y_1(x)$ in the interior 
of the sector $\arg x\in[\frac{3\pi}{5},\pi]$, we apply (\ref{y+_as}). 
Similarly, we find the asymptotics of the solution $y_0(x)$ in the 
interior of the sector $\arg x\in[\pi,\frac{7\pi}{5}]$ using 
(\ref{y-_as}). Either expression implies
\begin{cor}\label{corollary1}
If $|x|\to\infty$, $\arg x\in[\frac{3\pi}{5},\frac{7\pi}{5}]$, then
\begin{equation}\label{y0_hat_y0}
y_1(x)-y_0(x)= \frac{i}{\sqrt\pi}2^{-11/8}3^{-1/8}
(e^{-i\pi}x)^{-1/8}e^{-\frac{1}{5}2^{11/4}3^{1/4}(e^{-i\pi}x)^{5/4}}
(1+{\cal O}(x^{-3/8})).
\end{equation}
\end{cor}

Applying symmetries (\ref{P_symmetries}) to $y_k(x)$, $k=0,1$, we find
the solutions $y_k(x)$ corresponding to the Stokes multipliers
$s_k=s_{k-1}=0$,
\begin{align}\label{yk_def}
y_{2n}(x)&=e^{i\frac{4\pi}{5}n}y_0(e^{i\frac{2\pi}{5}n}x)\quad
\text{for}\quad s_{2n}=s_{2n-1}=0,\notag \\
y_{2n+1}(x)&=e^{i\frac{4\pi}{5}n}y_1(e^{i\frac{2\pi}{5}n}x)\quad
\text{for}\quad s_{2n+1}=s_{2n}=0.
\end{align}
Since there is one-to-one correspondence between the points of the
monodromy surface and the Painlev\'e functions, the first of the
identities (\ref{scalar_relations}), $s_{k+5}=s_k$, implies that
$y_{n+5}(x)=y_n(x)$.

Using Theorems~\ref{Theorem2} and~\ref{Theorem3}, we find that
\begin{align}\label{y4n}
y_{4n}(x)=y_{4n+5}(x)&=\sqrt{e^{-i\pi}\tfrac{x}{6}} +{\cal O}(x^{-2}),
\\ &\quad |x|\to\infty,\quad \arg x\in
[\tfrac{\pi}{5}-\tfrac{4\pi}{5}n,\pi-\tfrac{4\pi}{5}n], \notag
\\\label{y4n-2} y_{4n-2}(x)=y_{4n+3}(x)&=-\sqrt{e^{-i\pi}\tfrac{x}{6}}
+{\cal O}(x^{-2}), \\ &\quad |x|\to\infty,\quad \arg x\in
[\tfrac{3\pi}{5}-\tfrac{4\pi}{5}n,\tfrac{7\pi}{5}-\tfrac{4\pi}{5}n].
\notag
\end{align}

The symmetry (\ref{P_symmetries}) with the definition (\ref{yk_def})
applied to (\ref{y0_hat_y0}) yields
\begin{cor}\label{corollary2}
If\/ $|x|\to\infty$\/ and\/ $\arg x\in
[\tfrac{3\pi}{5}-\tfrac{2\pi}{5}n,\tfrac{7\pi}{5}-\tfrac{2\pi}{5}n]$,
then
\begin{multline}\label{yk_hat_yk}
y_{2n+1}(x)-y_{2n}(x)= \\
=\frac{e^{i\frac{\pi}{2}+i\frac{4\pi}{5}n}}{\sqrt\pi}2^{-11/8}3^{-1/8}
(e^{-i\pi+i\frac{2\pi}{5}n}x)^{-1/8}
e^{-\frac{1}{5}2^{11/4}3^{1/4}(e^{-i\pi+i\frac{2\pi}{5}n}x)^{5/4}}
\times \\ \times (1+{\cal O}(x^{-3/8})),
\end{multline}
\end{cor}
On the one hand, equations (\ref{y0_hat_y0}), (\ref{yk_hat_yk})
constitute the quasi-linear Stokes phenomenon for the Painlev\'e first
equation. On the other hand, these equations give the asymptotic
description of the degenerate Painlev\'e functions beyond the sectors
in (\ref{y4n}) and (\ref{y4n-2}). Observing that the difference
(\ref{yk_hat_yk}) is exponentially small in the interior of the
indicated sector, we conclude that the asymptotics (\ref{y4n}) and
(\ref{y4n-2}) as $|x|\to\infty$ continue to wider open sectors,
\begin{align}\label{y4n_ext}
y_{4n}(x)&=\sqrt{e^{-i\pi}\tfrac{x}{6}}+{\cal O}(x^{-2}),\quad \arg
x\in (\epsilon-\tfrac{\pi}{5}-\tfrac{4\pi}{5}n,
\tfrac{7\pi}{5}-\tfrac{4\pi}{5}n-\epsilon), \\\label{y4n-2_ext}
y_{4n-2}(x)&=-\sqrt{e^{-i\pi}\tfrac{x}{6}}+{\cal O}(x^{-2}),\quad \arg
x\in (\epsilon+\tfrac{\pi}{5}-\tfrac{4\pi}{5}n,
\tfrac{9\pi}{5}-\tfrac{4\pi}{5}n-\epsilon),
\end{align}
where $\epsilon>0$ is an arbitrary small constant.
\begin{remark}\label{rem3}
The solutions $y_n(x)$ (\ref{yk_def}) corresponding to the trivial
values of two Stokes multipliers $s_n=s_{n-1}=0$ are the most
degenerate among the Painlev\'e transcendent since they behave
algebraically in four of five sectors $\arg x\in
(-\frac{\pi}{5}+\frac{2\pi}{5}k,\frac{\pi}{5}+\frac{2\pi}{5}k)$,
$k=0,\pm1,\pm2$, see (\ref{y4n_ext}), (\ref{y4n-2_ext}). Nevertheless
these solutions are transcendent, since their asymptotics as 
$|x|\to\infty$ within the remaining fifth sector involves the 
elliptic function of Weierstra\ss, look for more details in 
\cite{kap_kit}. Moreover, the fact that the asymptotics of $y_n(x)$ 
is not elliptic in four sectors uniquely determines the values of 
all the Stokes multipliers $s_k$. Thus the asymptotics 
(\ref{y4n_ext}), (\ref{y4n-2_ext}) uniquely determine the 
degenerate solutions $y_n(x)$.
\end{remark}
\begin{remark}\label{rem4}
The asymptotics of less degenerate solutions corresponding to
$s_n=0$ and $s_{n+1}+s_{n-1}=i$ can be found applying the symmetries
(\ref{P_symmetries_rotate}) to equations (\ref{y+_as}) and
(\ref{y-_as}).
\end{remark}

\section{Coefficient asymptotics}\label{as_coeffs}

Using the steepest descent approach, cf.\ \cite{DZ2}, we can show the
existence of the asymptotic expansion of $y_n(x)$, $n\in{\Bbb Z}$, in
the negative degrees of $x^{1/2}$. Further elementary investigation of
the recursion relation for the coefficients of the series allows us to
claim that the asymptotic expansion for $y_n(x)$ in (\ref{y4n_ext}),
(\ref{y4n-2}) has the following form:
\begin{multline}\label{y+formal}
y_f(x)=\sigma\bigl(-\tfrac{x}{6}\bigr)^{1/2}
\sum_{k=0}^{\infty}a_k\sigma^k(-x)^{-5k/2} +{\cal O}(x^{-\infty})= \\
=\sigma\bigl(-\tfrac{x}{6}\bigr)^{1/2}
\sum_{k=0}^{\infty}a_{2k}(-x)^{-5k} +\tfrac{1}{\sqrt6}(-x)^{-2}
\sum_{k=0}^{\infty}a_{2k+1}(-x)^{-5k} +{\cal O}(x^{-\infty}),\quad
\sigma^2=1,
\end{multline}
where coefficients $a_k$ are determined uniquely by the recurrence
relation
\begin{equation}\label{an_recurrence}
a_0=1,\quad a_{k+1}=\tfrac{25k^2-1}{8\sqrt6}a_k
-\tfrac{1}{2}\sum_{m=1}^ka_ma_{k+1-m}.
\end{equation}
Several initial terms of the expansion are given by
\begin{multline}\label{y0_expansion}
y_f(x)=\sigma\sqrt{-x/6}\Bigl\{ 1+\tfrac{49}{768x^5}
-\tfrac{4412401}{1179648x^{10}} +\tfrac{245229441961}{100663296x^{15}}
+{\cal O}(x^{-20})\Bigr\}- \\ -\tfrac{1}{48x^2}\Bigl\{
1-\tfrac{1225}{192x^5} +\tfrac{73560025}{49152x^{10}}
-\tfrac{7759635184525}{3538944x^{15}} +{\cal O}(x^{-20})\Bigr\}.
\end{multline}

Our next goal is to determine the asymptotics of the coefficients
$a_k$ in (\ref{y+formal}) as $k\to\infty$. With this purpose, let us
construct a sectorial analytic function $\hat y(t)$,
\begin{equation}\label{hat_y}
\arg t\in [-\tfrac{2\pi}{5}(n+1),-\tfrac{2\pi}{5}n]\colon \quad \hat
y(t)=y_{4n}(e^{i\pi}t^2),\quad n=-2,-1,0,1,2.
\end{equation}
The function $\hat y(t)$ has a finite number of poles all contained 
in a circle $|t|<\rho$ and is characterized by the uniform asymptotic 
expansion near infinity,
\begin{equation}\label{hat_y_expansion}
\hat y(t)=\frac{t}{\sqrt6} \sum_{k=0}^{\infty}a_kt^{-5k} +{\cal
O}(t^{-\infty}).
\end{equation}

Let $y^{(N)}(t)$ be a partial sum
\begin{equation}\label{yN}
y^{(N)}(t)=\frac{t}{\sqrt6} \sum_{k=0}^{N-1}a_kt^{-5k},
\end{equation}
and $v^{(N)}(t)$ be a product
\begin{equation}\label{vn}
v^{(N)}(t)=t^{5N-2}\sqrt{6}\bigl(\hat y(t)-y^{(N)}(t)\bigr)=
t^{-1}\sum_{k=0}^{\infty}a_{k+N}t^{-5k}+{\cal O}(t^{-\infty}).
\end{equation}
Because $t^{5N-2}y^{(N)}(t)$ is polynomial, the integral of
$v^{(N)}(t)$ along the counter-clock-wise oriented circle of the
radius $|t|=\rho$ satisfies the estimate
\begin{equation}\label{vN_integral}
\Bigl|\oint_{|t|=\rho}v^{(N)}(t)\,dt\Bigr|\leq
\rho^{5N-2}\sqrt{6}\oint_{|t|=\rho}|\hat y(t)|\,dl\leq
\sqrt{6}\,2\pi\rho^{5N-1}\max_{|t|=\rho}|\hat y(t)|=C\rho^{5N}
\end{equation}
with some positive constant $C$ whose precise value is not important
for us.

On the other hand, inflating the sectorial arcs of the circle
$|t|=\rho$, we find that
\begin{equation}\label{vN_inflation}
\oint_{|t|=\rho}v^{(N)}(t)\,dt=\oint_{|t|=R}v^{(N)}(t)\,dt
+\sum_{n=-2}^2\int_{e^{i\frac{2\pi}{5}n}(\rho,R)}
\bigl(v_+^{(N)}(t)-v_-^{(N)}(t)\bigr)\,dt.
\end{equation}
Because $v^{(N)}(t)=t^{-1}a_N+{\cal O}(t^{-6})$, the first of the
integrals in the r.h.s.\ of (\ref{vN_inflation}) is computed as
follows:
\begin{equation}\label{aN_int}
\oint_{|t|=R}v^{(N)}(t)\,dt=2\pi i a_N+{\cal O}(R^{-5}).
\end{equation}
Remaining integrals in (\ref{vN_inflation}) are computed using
definitions (\ref{hat_y})--(\ref{vn}) and (\ref{yk_def}) with the
identification $y_{-4}(x)=y_1(x)$ and the formula (\ref{y0_hat_y0})
together,
\begin{multline}\label{eval}
\sum_{n=-2}^2 \int\limits_{e^{i\frac{2\pi}{5}n}(\rho,R)} \!\!
\bigl(v_+^{(N)}(t)-v_-^{(N)}(t)\bigr)\,dt=
5\sqrt6\int\limits_{(\rho,R)}t^{5N-2}
\bigl(y_{-4}(e^{i\pi}t^2)-y_0(e^{i\pi}t^2)\bigr)\,dt= \\
=i\frac{5\sqrt6}{\sqrt\pi}2^{-11/8}3^{-1/8} \int\limits_{(\rho,R)}
t^{5N-\frac{9}{4}} e^{-\frac{1}{5}2^{11/4}3^{1/4}t^{5/2}} (1+{\cal
O}(t^{-3/4}))\,dt= \\ =2i\frac{\sqrt6}{\sqrt5\sqrt\pi}
\bigl(\tfrac{1}{5}2^{11/4}3^{1/4}\bigr)^{-2N} \Gamma(2N-\tfrac{1}{2})
(1+{\cal O}(N^{-3/10})) +{\cal O}(\rho^{5N-\frac{5}{2}})+ \\ +{\cal
O}\bigl( e^{-\frac{1}{5}2^{11/4}3^{1/4}R^{5/2}}R^{5N-\frac{15}{4}}
\bigr).
\end{multline}
Thus, letting $R=\infty$, we find the asymptotics
of the coefficient $a_N$ in (\ref{y+formal}) as $N\to\infty$,
\begin{equation}\label{an_fin}
a_N=-\frac{\sqrt6}{\sqrt5\,\pi^{3/2}}
\bigl(\tfrac{1}{5}2^{11/4}3^{1/4}\bigr)^{-2N} \Gamma(2N-\tfrac{1}{2})
(1+{\cal O}(N^{-3/10})) +{\cal O}(\rho^{5N}),\quad N\to\infty.
\end{equation}
\begin{remark}\label{rem5}
The presented asymptotic formula shows a remarkable accuracy: neglecting 
in (\ref{an_fin}) error terms, we find an approximation to $a_N$ with 
the relative error not exceeding 2\% for $N=4$ and 1\% for $N=7$. 
Furthermore, for the initial set of $N=1,2,\dots,7$, the relative error 
decreases approximately as $N^{-1}$ which is significantly better than 
estimated.
\end{remark}

\bigskip
{\bf Acknowledgments.} This work was partially supported by the RFBR, 
grant No.~02--01--00268. The author is grateful to A.R.~Its for important 
remarks and suggestions.

\ifx\undefined\bysame \newcommand{\bysame}{\leavevmode\hbox
to3em{\hrulefill}\,} \fi


\begin{thebibliography}{References}

\bibitem{painleve1} 
P. Painlev\'e, Sur la d\'etermination explicite des \'equations 
diff\'erentielles du second ordre \`a points critiques fixes, 
{\it Comptes Rendus} {\bf 127} (1898) 945--948.

\bibitem{ince} 
E.L. Ince, {\it Ordinary Differential Equations} (New York: Dover), 1965.

\bibitem{painleve3} P. Painlev\'e, Sur les \'equations
diff\'erentielles du second ordre \`a points critiques fixes, {\it
Comptes Rendus} {\bf 143} (1906) 1111--1117.

\bibitem{abl_segur} M.J. Ablowitz and H. Segur, {\it Solitons and the
inverse scattering transform}, SIAM, Philadelphia, 1981.

\bibitem{TSB} D.L. Turcotte, D.A. Spence and H.H. Bau, {\i Int.\ J.\
Heat Mass Transfer}, {\bf 25} (1982) 699--706.

\bibitem{haberman} 
R. Haberman, Slowly varying jump and transition
phenomena associated with algebraic bifurcation problems, 
{\it J.\ Appl.\ Math.} {\bf 51} (1979) 69--106;
Slow passage through the nonhyperbolic homoclinic orbit associated 
with a subcritical pitchfork bifurcation for Hamiltonian systems and 
the change in action, {\it SIAM J.\ Appl.\ Math.} {\bf 62} (2001) no.~2, 
488-513.

\bibitem{DS} M. Douglas and S. Shenker, Strings in less than one
dimension, {\it Nucl.\ Phys.} B {\bf 335} (1990) 635-654.

\bibitem{GM} D. Gross and A. Migdal, A nonperturbative treatment of
two-dimensional quantum gravity, {\it Nucl.\ Phys.} B {\bf 340} (1990)
333-365.

\bibitem{FIK1} A.S. Fokas, A.R. Its, A.V. Kitaev, Discrete Painlev\'e
equations and their appearance in quantum gravity, {\it Comm.\ Math.\
Phys.} {\bf 142} (1991), no. 2, 313--344; The isomonodromy approach to
matrix models in $2$D quantum gravity, {\it Comm.\ Math.\ Phys.}  {\bf
147} (1992), no. 2, 395--430.

\bibitem{PS}
L. D. Paniak and R. J. Szabo, Fermionic quantum gravity, {\it Nucl.\ Phys.}
B {\bf 593} (2001) 671-725.

\bibitem{FIK2} A.S. Fokas, A.R. Its and A.V. Kitaev, Matrix models of
two-dimensional quantum gravity, and isomonodromic solutions of
Painlev\'e ``discrete equations'', {\it Zap.\ Nauchn.\ Sem.\ LOMI}
{\bf 187} (1991) 3--30; translation in: {\it J.\ Math.\ Sci.} {\bf 73}
(1995), no. 4, 415--429.

\bibitem{kapaev:sclim}
A. Kapaev, Monodromy approach to the scaling limits in 
isomonodromy systems, {\it Theor.\ Math.\ Phys.} {\bf 137}(3) (2003)
1691--1702; nlin.SI/0211022.

\bibitem{ABWZ} 
O. Agam, E. Bettelheim, P. Wiegmann and A. Zabrodin, Viscous fingering and a
shape of an electronic droplet in a quantum Hall regime,
{\it Phys.\ Rev.\ Lett.} {\bf 88} (2002) 236801; cond-mat/0111333. 

\bibitem{novikov}
S.P. Novikov, Quantization of finite-gap potentials and nonlinear
quasiclassical approximation in nonperturbative string theory, 
{\it Funct.\ Anal.\ Appl.} {\bf 24} (1990) 296--306.

\bibitem{FGMR}
F. Fucito, A. Gamba, M. Martellini and O. Ragnisco, Nonlinear WKB
analysis of the string equations, {\it Internat.\ J.\ Modern Phys.} B
{\bf 6} (1992) 2123--2148.

\bibitem{eynard_zinn-justin}
B. Eynard and J. Zinn-Justin, Large order behavior of 2D gravity coupled to
$D<1$ matter, {\it Phys.\ Lett.\ B} {\bf 302} (1993) 394-402.

\bibitem{DiFGZJ}
P. Di Francesco, P. Ginsparg and J. Zinn-Justin, 2D gravity and random
matrices, {\it Phys. Rep.} {\bf 254} (1995) no.~1-2, 1-133.

\bibitem{wiegmann}
P. Wiegmann and R. Teodorescu, private communication (2003).

\bibitem{boutroux} P. Boutroux, Recherches sur les transcendantes de
M. Painlev\'e et l'etude asymptotique des \'equations
diff\'erentielles du second ordre {\it Ann.\ Sci.\ Ecol.\ Norm.\
Sup\'er.} {\bf 30} (1913) 255--376; {\it Ann.\ Sci.\ Ecol.\ Norm.\
Sup\'er.} {\bf 31} (1914) 99--159.

\bibitem{BO}
C.M. Bender and S.A. Orszag, {\it Advanced Mathematical methods for
scientists and engineers}, McGraw Hill, New York, 1978.

\bibitem{HS}
P. Holmes and D. Spence, On a Painlev\'e-type boundary-value problem,
{\it Quart.\ J.\ Mech.\ Appl.\ Math.} {\bf 37} (1984) 525--538.

\bibitem{hille}
E. Hille, {\it Lectures on ordinary differential equations}, Addison
Wesley, Reading, Mass., 1986.

\bibitem{JK}
N. Joshi and A. Kitaev, On Boutroux tritronqu\'ee solutions of the first
Painlev\'e equation {\it Stud.\ Appl.\ Math.} {\bf 107} (2001) 253--291.

\bibitem{joshi_kruskal}
N. Joshi and M.D. Kruskal, An asymptotic approach to the connection
problem for the first and the second Painlev\'e equations, 
{\it Phys.\ Lett.\ A} {\bf 130} (1988) 129--137.

\bibitem{joshi_kruskal2}
N. Joshi and M.D. Kruskal, The Painlev\'e connection problem: an asymptotic
approach. I, {\it Stud.\ Appl. Math.} {\bf 86} (1992) 315--376.

\bibitem{jm2}
M. Jimbo, T. Miwa and K. Ueno, Monodromy preserving deformation of
linear ordinary differential equations with rational coefficients
{\it Physica D} {\bf 2} (1981) 306--352\hfil\par
M. Jimbo and T. Miwa, Monodromy preserving deformation of linear
ordinary differential equations with rational coefficients. II 
{\it Physica D} {\bf 2} (1981) 407--448;\hfil\par
\bysame Monodromy preserving deformation of linear ordinary
differential eq\-u\-a\-ti\-ons with rational coefficients. III 
{\it Physica D} {\bf 4} (1981) 26--46.

\bibitem{FN} 
H. Flaschka and A.C. Newell, Monodromy- and spectrum-preserving 
deformations I, {\it Comm.\ Math.\ Phys.} {\bf 76} (1980) 65-116.

\bibitem{its_nov}
A.R. Its and V.Yu. Novokshenov, The Isomonodromic Deformation Method
in the Theory of Painlev\'e Equations, {\it Lect.\ Notes Math.} 
{\bf 1191}, 1-313, Berlin-Heidelberg-New York-Tokyo: Springer-Verlag,
1986.

\bibitem{kapaevP1}
A.A. Kapaev, Asymptotics of solutions of the Painlev\'e equation of the
first kind {\it Diff.\ Eqns.} {\bf 24} (1989) 1107--1115 (translated from:
{\it Diff.\ Uravnenija} {\bf 24} (1988) 1684--1695 (Russian)).

\bibitem{kap_kit}
A.A. Kapaev and A.V. Kitaev, Connection formulae for the first Painlev\'e
transcendent in the complex domain {\it Lett.\ Math.\ Phys.} {\bf 27}
(1993) 243--252.

\bibitem{kapaev:scl2}
A. Kapaev, Monodromy deformation approach to the scaling limit of the
Painlev\'e first equation, {\it CRM Proc.\ Lect.\ Not.} {\bf 32}
(2002) 157--179; nlin.SI/0105002.

\bibitem{takei}
Y. Takei, On the connection formula for the first Painlev\'e equation --
from the viewpoint of the exact WKB analysis,
{\it S\=urikaisekiken\-ky\=usho K\=oky\=uroku} {\bf 931} (1995)
70--99. 

\bibitem{its_kap}
A.R. Its and A.A. Kapaev, Quasi-linear Stokes phenomenon for the
second Painlev\'e transcendent, {\it Nonlinearity} {\bf 16} (2003)
363--386; nlin.SI/0108010.

\bibitem{DZ}
P.A. Deift and X. Zhou, A steepest descent method for oscillatory
Riemann-Hilbert problems. Asymptotics for the MKdV equation
{\it Ann.\ of Math.} {\bf 137} (1995) 295--368.

\bibitem{HJK}
A. N. W. Hone, N. Joshi and A. V. Kitaev, An entire function defined by a
nonlinear recurrence relation, {\it J.\ London Math.\ Soc.} {\bf 66}(2)
(2002) 377--386.

\bibitem{AK}
F.V. Andreev and A.V. Kitaev, Exponentially small corrections to divergent
asymptotic expansions of solutions of the fifth Painlev\'e equation
{\it Math.\ Res.\ Lett.} {\bf 4} (1997) 741--759.

\bibitem{garnier1} R.~Garnier, Sur les \'equations diff\'erentielles
du troisi\`me ordre dont l'int\'egrale g\'en\'erale est uniforme et
sur une classe d'\`quations nouvelles d'ordre sup\'erieur dont
l'int\'egrale g\'en\'erale a ses points critique fixes, {\it
Ann. Sci. \'Ecole Norm. Sup.} (4) {\bf 29} (1912), 1--126.

\bibitem{garnier3} R.~Garnier, Solution du probl\`me de Riemann pour
les syst\`mes diff\'erentielles lin\'eaires du second ordre, {\it
Ann.\ Sci.\ \'Ecole Norm.\ Sup.} {\bf 43} (1926) 177--307.
  
\bibitem{wasow}
W.~Wasow, {\it Asymptotic Expansions for Ordinary Differential 
Equations}, Intersci\-ence-Wiley, New York, 1965.

\bibitem{BE}
H.~Bateman and A.~Erdelyi, {\it Higher Transcendental Functions}, 
McGraw-Hill, New York, 1953.

\bibitem{olver}
F.~W.~J.~Olver, {\it Asymptotics and special functions}, Academic Press,
New York, 1974.

\bibitem{DZ2}
P. Deift and X. Zhou, Long-time asymptotics for integrable systems. Higher
order theory {\it Comm.\ Math.\ Phys.} {\bf 165} (1995) 175--191.

\end{thebibliography}
\end{document}